\documentclass[onecolumn]{aa}  
\usepackage{natbib,,twoopt}
\usepackage[breaklinks=true]{hyperref} 
\bibpunct{(}{)}{;}{a}{}{,}             
\makeatletter
  \newcommandtwoopt{\citeads}[3][][]{\href{http://adsabs.harvard.edu/abs/#3}%
    {\def\hyper@linkstart##1##2{}%
     \let\hyper@linkend\@empty\citealp[#1][#2]{#3}}}
  \newcommandtwoopt{\citepads}[3][][]{\href{http://adsabs.harvard.edu/abs/#3}%
    {\def\hyper@linkstart##1##2{}%
     \let\hyper@linkend\@empty\citep[#1][#2]{#3}}}
  \newcommandtwoopt{\citetads}[3][][]{\href{http://adsabs.harvard.edu/abs/#3}%
    {\def\hyper@linkstart##1##2{}%
     \let\hyper@linkend\@empty\citet[#1][#2]{#3}}}
  \newcommandtwoopt{\citeyearads}[3][][]%
    {\href{http://adsabs.harvard.edu/abs/#3}
    {\def\hyper@linkstart##1##2{}%
     \let\hyper@linkend\@empty\citeyear[#1][#2]{#3}}}
\makeatother
\usepackage{graphicx}
\usepackage{txfonts}
\usepackage{color}

\begin{document}

   \title{Extended Skyrme Equation of State in asymmetric nuclear matter }


   \author{          D. Davesne \inst{1}  \and A. Pastore \inst{2} \and J. Navarro \inst{3}
         }

   \institute{
             Universit\'e Lyon 1, Institut de Physique Nucl{\'e}aire de Lyon, UMR 5822, CNRS-IN2P3, 
             43 Bd. du 11 Novembre 1918, F-69622 Villeurbanne cedex, France
             \and CEA, DAM, DIF, F-91297 Arpajon, France
             \and IFIC (CSIC-Universidad de Valencia), Apartado Postal 22085, E-46.071-Valencia, Spain
             }

   \date{8th June 2015}

  \abstract{We present a new equation of state for infinite systems (symmetric, asymmetric and neutron matter) based on an \emph{extended} Skyrme functional constrained by microscopic Brueckner-Bethe-Goldstone  results. The resulting equation of state reproduces with very good accuracy the main features of microscopic calculations and it is compatible with recent measurements of two times Solar-mass neutron stars. We provide all necessary analytical expressions to facilitate a quick numerical implementation of quantities of astrophysical interest.}


\maketitle
%

\section{Introduction}

A key ingredient for many astrophysical calculations is a reliable Equation of State (EoS) for isospin asymmetric matter, covering from symmetric nuclear matter (SNM) to pure neutron matter (NM), from low to high densities ($\simeq 4-5$ times saturation density). In this respect, a popular application is by instance the description of neutron stars (NS) properties as the mass-radius relation or their inhomogeneous crust.
Restricting ourselves to the category of nucleonic EoS, one of the most popular EoS is the one derived by ~\cite{bal97}. It has been obtained within the context of Brueckner-Bethe-Goldstone (BBG) many-body theory using the Argonne v14 potential plus the Urbana model for the three-body nuclear interaction.
Such an EoS has been tabulated for given values of the density of the system.
For such a reason, it is customary to fit the EoS with some analytical expressions which are much simple to handle in numerical codes (\cite{typ13}) suited for astrophysical simulations. 

A possible alternative to the fit is the use of an effective Skyrme interaction (\cite{sky59}) as early suggested by~\cite{cao06}. These authors have shown that it is possible to fit an effective Skyrme functional (that they named LNS) on BBG, conserving some of the main features of the original BBG EoS.
The main advantage of using a functional instead of a generic  interpolation, as done for instance by \cite{hae04}, is that once the parameters of the functional are fixed, all basic properties as pression or symmetry energy can be simply obtained by standard derivative operations. In the case where the vector part of the functional is also taken into account, as for the case of a functional derived from a complete Skyrme interaction, with a simple formalism based on the Linear Response theory~\cite{PRep}, one can also describe collective phenomena within the NS. The latter play a crucial role in describing different phenomena in the NS as the thermal properties of the inner crust (\cite{cha13}) or the neutrino mean free path~\cite{iwa82}.
Another important advantage is that  the same functional can also be consistently used to describe the region of the crust of the NS (\cite{cha08}), thus allowing for a unified description of the star. Although LNS gives a nice reproduction on infinite matter properties up to two times saturation density, its EoS in pure neutron matter (PNM) remarkably deviates from the BBG results, leading to a different behavior of the symmetry energy at high density. It follows that the LNS EoS supports only NS with mass lower than 1.6 Solar-masses as shown by  \cite{Sin13}. 
The authors of (\cite{gam11}) have recently refitted the LNS functional, but hte new LNS1 and LNS5 do not substantially improve the properties of the homogeneous nuclear medium as compared to the original LNS, although they improve the description of finite nuclei.
In the present article, we generalize the analysis done by \cite{cao06} concerning the possibility of constraining a phenomenological Skyrme functional on microscopic results, by using an \emph{extended} Skyrme functional which includes up to 6th order derivative terms (\cite{car08,rai11}), aiming at giving a reliable EoS also in the high density region and thus in better agreement with BBG results.

Indeed, the Skyrme interaction can be interpreted as a low-momentum expansion of a finite-range interaction (\cite{sky59}). The \emph{standard} form of the interaction is the one given by ~\cite{vau72} and it takes into account only gradients terms up to the second power, as for LNS.
Although this can be viewed as a good approximation to be used in finite nuclei calculations (\cite{ben03}), it is not adapted to the study of dense nuclear matter. For example, the \emph{standard} Skyrme interaction is not able to reproduce at the same time the correct isovector splitting of the effective mass of BBG results (\cite{bal14}) and the high density behavior of nuclear matter. Among the different strategies one can adopt to overcome these difficulties, the most promising implies either the addition of extra density dependencies on the velocity dependent terms (\cite{cha10}) or the inclusion of higher order derivative terms  (\cite{car08}). We prefer to follow the latter approach since it allows to grasp the correct behavior of the EoS of BBG calculations especially at high density. In particular, it has been shown (\cite{dav15})  that the different terms follow a precise hierarchy and thus high order terms have stronger influence on the high density part, inducing almost negligible modification to the low density part. Such a result is also in good agreement with previous findings of \cite{car10}, based on Density Matrix Expansion methods in finite nuclei. To this respect, the functional presented here can be considered as the natural extension of the LNS one to correct the high density region.
In the present article, we thus present an \emph{extended} Skyrme functional, hereafter called LYVA1, for a proper treatment of these higher order gradients, giving all necessary analytical expressions for astrophysical calculations. A numerical code and the tabulated values of this new EoS will be available at the CompOSE webpage\footnote{http://compose.obspm.fr/}.

The article is organized as follows. In Sec.~\ref{Sec:Sky} we present the general formalism of the extended Skyrme functional. In Sec.~\ref{Sec:BA} we give the general formula for the binding energy per particle for isospin asymmetric nuclear matter, while in Sec.~\ref{Sec:BC} we present the case of polarized matter. In Sec.~\ref{Sec:esym}, we study the behavior of the symmetry energy. In Sec.\ref{Sec:effmass} we further discuss the behavior of the effective mass and in Sec.~\ref{sec:ns}  we examine the applications of our model to the description of a NS. Our conclusions are then given in Sec.\ref{sec:concl}.

%
%
%
%
\section{Extended Skyrme interaction}\label{Sec:Sky}

 The most general form of the Skyrme functional up to 6th order in the gradient expansion has been derived by \cite{car08}. In the present article, we prefer to relate this functional to an effective interaction, thus reducing the number of free coupling constants, as shown in (\cite{dav13}). The corresponding Skyrme interaction reads~(\cite{rai11,dav14c})

\begin{eqnarray}\label{sk:ext}
v^{}_{Sk}&=&t^{(0)}_{0}\left(1+x^{(0)}_{0}P_{\sigma}\right)+\frac{1}{2}t^{(2)}_{1}\left(1+x^{(2)}_{1}P_{\sigma}\right)\left[\mathbf{k}^{'2}+\mathbf{k}^{2} \right]+t^{(2)}_{2}\left(1+x^{(2)}_{2}P_{\sigma}\right)\mathbf{k}^{'}\cdot \mathbf{k}+\frac{1}{6}t^{(0)}_{3}\left(1+x^{(0)}_{3}P_{\sigma}\right)n^{\alpha}(\mathbf{R})\nonumber\\
&+&\frac{1}{4}t_{1}^{(4)}\left(1+x^{(4)}_{1}P_{\sigma}\right)\left[(\mathbf{k}^{'2}+\mathbf{k}^{2})^{2}+4(\mathbf{k}'\cdot\mathbf{k})^{2} \right]+t_{2}^{(4)}\left(1+x^{(4)}_{2}P_{\sigma}\right)(\mathbf{k}^{'}\cdot \mathbf{k})(\mathbf{k}^{'2}+\mathbf{k}^{2})\,\nonumber\\
&+&\frac{1}{2}t_{1}^{(6)}\left(1+x^{(6)}_{1}P_{\sigma}\right)(\mathbf{k}^{'2}+\mathbf{k}^{2})\left[(\mathbf{k}^{'2}+\mathbf{k}^{2})^{2}+12(\mathbf{k}'\cdot\mathbf{k})^{2} \right]+t_{2}^{(6)}\left(1+x^{(6)}_{2}P_{\sigma}\right)(\mathbf{k}^{'}\cdot \mathbf{k})\left[ 3(\mathbf{k}^{'2}+\mathbf{k}^{2})^{2}+4(\mathbf{k}^{'}\cdot\mathbf{k})^{2}\right]\, .
\end{eqnarray}

The notations used here are standard and more details can be found in ~(\cite{ben03}). The spin-orbit and tensor terms  are here discarded since they do not contribute to the total EoS, although they do contribute to its multipolar partial wave decomposition, as shown by  \cite{dav14p}. 
 The corresponding functional form can be obtained by performing an average on Hartree-Fock states. Results for an homogeneous medium are given in Eq.~(\ref{eosanm}).
An important advantage of deriving a functional from an effective interaction is that when applied to a single nucleon the functional leads to  vanishing internal energy. This is not automatically guarantee for a generic phenomenological functional  thus implying that the nucleon can interacting with itself~(\cite{cha10B}).

The parameters $t^{(n)}_{i}, x^{(n)}_{i}, (n=0,2,4,6)$ of this effective interaction have been fitted, following the method illustrated by ~\cite{dav15}, to some results of a BBG calculation (\cite{bal97,bal14a}), based on the microscopic Argonne $v14$ nucleon-nucleon two-body interaction plus the Urbana model for the three-body term. 
Although more recent BBG calculations for the EoS are available (as for instance \cite{zho04}), the complete determination of the parameters requires also the $ST$-decomposition of the potential energy. To the best of our knowledge, the BBG results of (\cite{bal97}) are the most complete, also including results for other important astrophysical quantities.  
There exist other {\em ab initio} results obtained from chiral effective field calculations at $V_{\text{low}-k}$ (\cite{heb11}) or the many-body perturbation theory (\cite{bog05,rot08}), which could also be used to fix extended Skyrme parameters with a similar quality (\cite{dav15}). However, they presently cover a narrower density range than BBG results, and are therefore not yet suited for our present purpose. We thus rely on the BBG results of (\cite{bal97}). The inputs for the fit  include the projection of the energy per particle in the  
different spin ($S$) and isospin ($T$) channels in symmetric nuclear matter (SNM), and the EoS of both SNM and pure neutron matter (PNM). As discussed by \cite{dav15}, no density-dependent term (i.e. $t^{(0)}_3, x^{(0)}_3, \alpha$ parameters) is required to get satisfactory fits, but the resulting parameterizations give too low a value for the Landau effective mass of SNM at saturation ($m^*/m \simeq 0.4$). In this paper, the density-dependent term is thus taken into account, and we have fixed its parameters to the values 
 $\alpha=1/6$, $t^{(0)}_{3}=13763$ [MeV\,fm$^{3+\alpha}$], and $x^{(0)}_{3}=0.3$.  
In such a way, we can properly constrain the higher order derivative terms whose role is mainly to give the correct asymptotic behavior at high density.

\begin{table}[h]
\begin{center}
\caption{Parameters of the extended LYVA1 Skyrme interaction,  
with $\alpha=1/6$, $t^{(0)}_{3}=13763$ [MeVfm$^{3+\alpha}$], and $x^{(0)}_{3}=0.3$.}
\begin{tabular}{ccccccc}
\hline
\hline
& & & & & &  \\
  $t^{(0)}_{0}$ [MeVfm$^{5}$] &  $t^{(2)}_{1}$ [MeVfm$^{5}$] &  $t^{(2)}_{2}$ [MeVfm$^{5}$]  &  $t_{1}^{(4)}$ [MeVfm$^{7}$] &$t_{2}^{(4)}$ [MeVfm$^{7}$] & $t_{1}^{(6)}$ [MeVfm$^{9}$] &$t_{2}^{(6)}$ [MeVfm$^{9}$] \\[2mm]
-2518.240 & 207.300& 527.930&  -23.691& -68.263 & 0&0.690 \\ [2mm]
\hline
& & & & & &  \\
    $x^{(0)}_{0}$  &  $x^{(2)}_{1}$ &  $x^{(2)}_{2}$   &  $x_{1}^{(4)}$ &$x_{2}^{(4)}$  & $x_{1}^{(6)}$ &$x_{2}^{(6)}$ \\[2mm]
0.2537 & -0.1688 & -1.0131 &  0.5650& -1.2022 & 0 & -1.2500 \\[2mm]
  \hline
 \hline
\end{tabular}
\label{tab:n3lo}
\end{center}
\end{table}

To fix the remaining parameters we have proceeded in two steps. In a first step, we considered the interaction (\ref{sk:ext}) up to 4th order in the gradient expansion only. The parameters $t^{(n)}_{i}, x^{(n)}_{i}, (n=0,2,4)$ have been determined by fitting BBG results for SNM EoS and $(S,T)$ channels. However, since the resulting EoS for PNM is too repulsive at high densities, we have added 6th order parameters on top of the previously determined 4th order values. To keep the quality of the SNM EoS we have imposed the values $t^{(6)}_{1}=0$, $x^{(6)}_{1}=0$, and $x^{(6)}_{2}=-5/4$. The remaining 6th order parameter  $t^{(6)}_2$ is then determined by fitting the PNM EoS in the full density interval. The resulting parameters of the LYVA1 interaction are given in Table \ref{tab:n3lo}.

\begin{figure*}[h]
   \centering
      \includegraphics[angle=-90,width=0.4\textwidth]{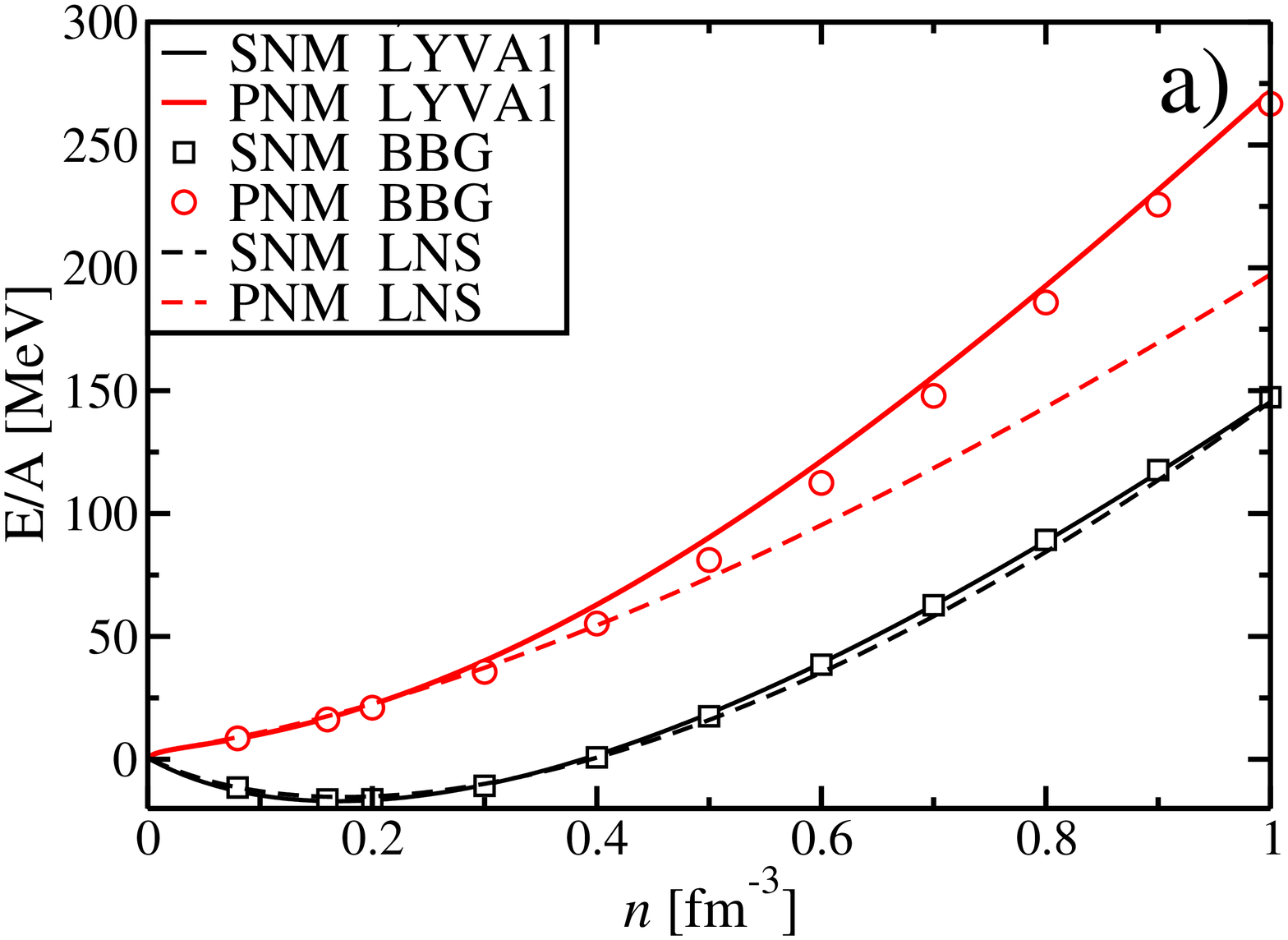}
   \includegraphics[angle=-90,width=0.4\textwidth]{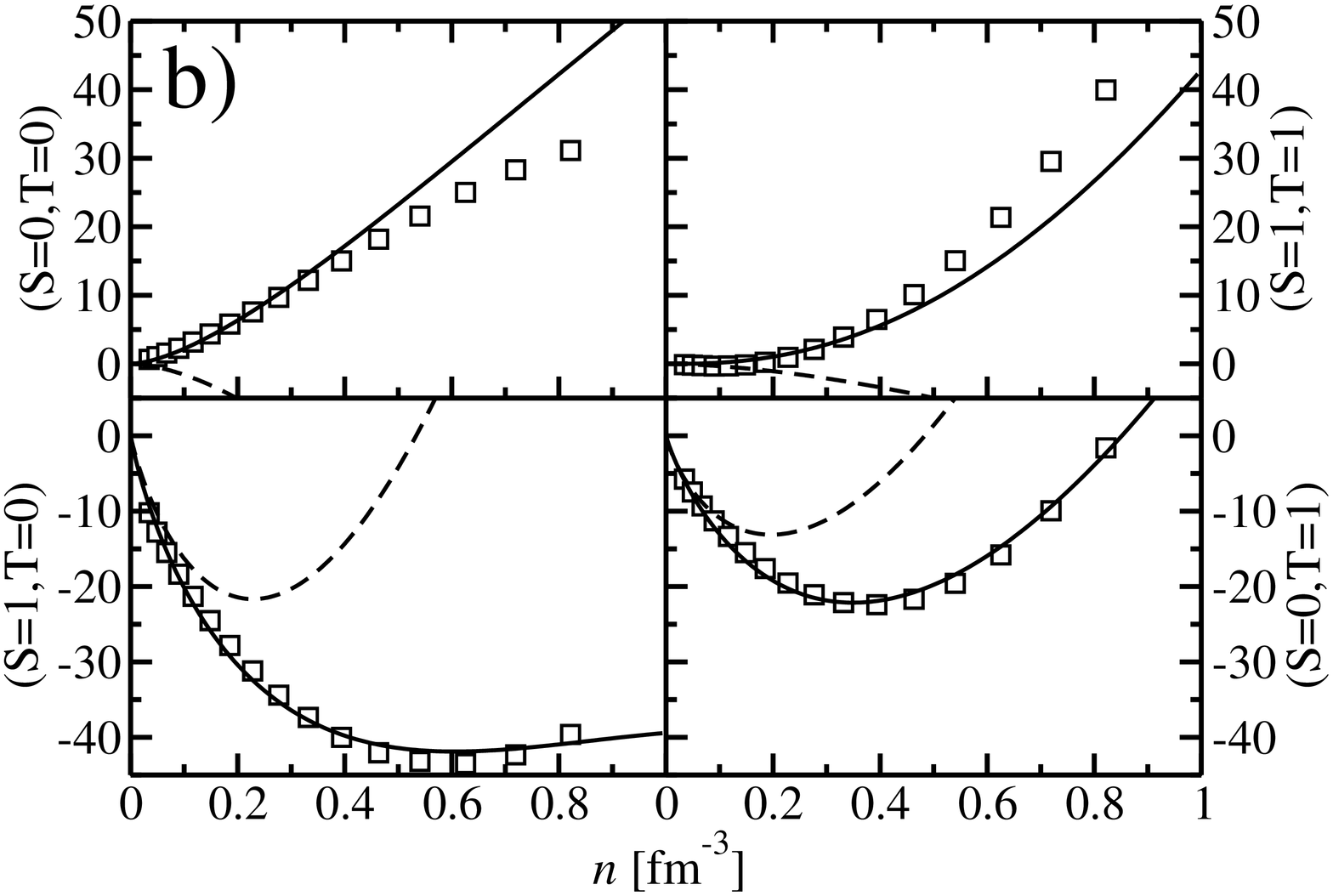}
   \caption{(Colors online) Equations of state of SNM and PNM (panel a) and  projections $(S,T)$ in SNM (panel b), both expressed in MeV. The solid lines represent the result obtained with our extended Skyrme interaction, while the dots represents the EoS obtained by \cite{bal97}. The LNS results are represented by dashed-lines.}
              \label{RESFIT}%
    \end{figure*}

In Figure \ref{RESFIT}, are displayed the EoS for SNM and PNM (panel a), and the SNM potential energy decomposition in the different spin-isospin ($S,T$) channels (panel b). Our fit is clearly very good for both EoS. The results obtained using LNS are also displayed (dashed lines), and one can see a rapid deviation of the PNM EoS, starting from $n\approx0.4$ fm$^{-3}$; one can expect this deviation also manifests for other quantities as the symmetry energy at high values of the density. Let us now turn to the results for the ($S,T$)-channels shown in panel b. As already discussed in (\cite{les06}), a general drawback of the \emph{standard} Skyrme functional is that the simultaneous reproduction of the ($S,T$)-channels is very difficult, to say the least. In the figure one can see the particular LNS case, which fails to reproduce BBG results. In contrast, with the extended functional the $S=1$ channels are nicely fitted in the full range of density values, whereas the $S=0$ channels show a deviation for $n \ge 0.6$ fm$^{-3}$ as a consequence of our giving more weigh to PNM data in the fit. All the other quantities presented hereafter in the article have not been fitted, and they can be considered as a prediction of our model.
 
\begin{figure*}[h]
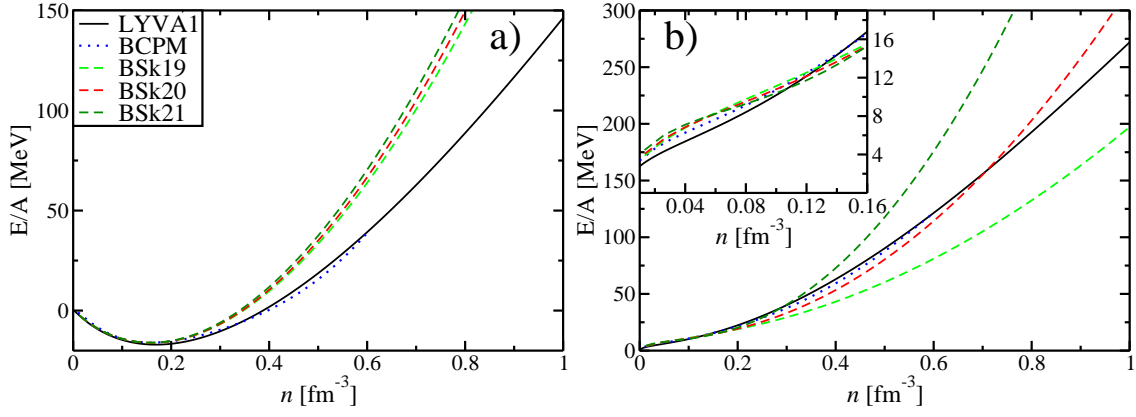

   \centering
      \includegraphics[angle=0,width=0.4\textwidth]{plots/EsurAsnmCOMPARE.eps}
      \includegraphics[angle=0,width=0.4\textwidth]{plots/EsurApnmCOMPARE.eps}
   \caption{(Colors online) Equations of state of SNM (panel a) and   PNM (panel b) obtained using the LYVA1 interaction (solid) the BSk models (dashed) and the BCMP functional (dotted).}
              \label{EoScompare}%
    \end{figure*}

It is worth mentioning that there are some other functionals which have been developed with particular attention to the properties of NS. Among the non-relativistic ones, we consider the BCPM~(\cite{bal13,bal14NS,sha15}) and the BSk family~(\cite{gor09,gor13}).
The BCPM functional has been derived in a complete Khon-Sham scheme, thus not related to any interaction, and it has been explicitly constrained to reproduce BBG results in homogeneous matter.
The BSk models have been derived from an effective Skyrme interaction with the addition of a power of the density into the momentum dependent terms of the standard Skyrme interaction. The BSk model have been constrained on several nuclear observables as masses and radii of finite nuclei together with additional pseudo-observables of homogeneous nuclear matter.  In Fig.\ref{EoScompare}, we compare the EoS in both SNM and PNM obtained with LYVA1, the BCPM functional and three representative BSk interactions, namely BSk19, BSk20 and BSk21~(\cite{cha11}).
Since the BCPM functional fits the same microscopic EoS as LYVA1, we observe that the results are almost on top of each other, except in the saturation region where the BCPM has been adjusted to give a value of saturation density $n_0=0.16$fm$^{-3}$ and $E/A=-16$~MeV. We remind that the BCMP has been fitted up to $n=0.6$~fm$^{-3}$ and that beyond that value the microscopic results of \cite{bur10} have been used. In order to be consistent, we will thus omit the points beyond $0.6$~fm$^{-3}$ in this paper.

\begin{figure}[h]
   \centering
   \includegraphics[angle=0,width=0.5\textwidth]{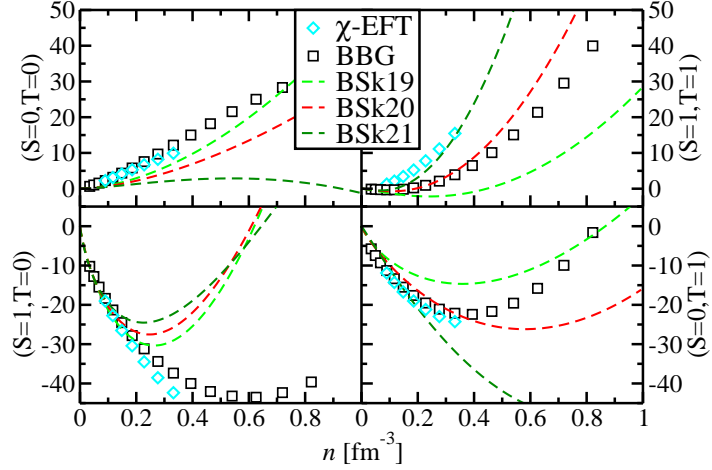}
   \caption{(Colors online) Same as Fig.\ref{RESFIT}, but for the BSk model discussed in the text.}
              \label{RESFIT2}%
    \end{figure}

Concerning the BSk functionals, we used the generalized expressions given in (\cite{les07}) to obtain their $ST$ decomposition. It is important to notice that as for the SLy4 case (~\cite{cha97}), the coupling constants in front of the so-called $J^{2}$ term are switched to zero.
This choice is justified in (~\cite{cha10}) to avoid the appearance of spurious ferromagnetic phase-transitions in the homogeneous medium~(\cite{mar02}) and anomalous behavior of the entropy.
In Fig.\ref{RESFIT2}, we compare the results obtained with the BSk models and the BBG calculations. We observe that the BSk behave better than any standard Skyrme interactions (\cite{les07}) since in the low-density region ($\approx n_{0}$) the BSk give the correct sign and trend of the energy per particle.
On the same figure we also report the chiral effective field theory ($\chi$-EFT) calculations at low momentum $k$ (\cite{heb11}). The ($\chi$-EFT) results are in very good agreement with the BBG results apart from the (S=1,T=1) channel. Such a comparison gives us the level of uncertainty related to the adopted interaction and/or calculation technique.

\section{Energy per particle }\label{Sec:BA}

In this section we give the analytical expression of the binding energy per particle ($E/A$) for infinite systems with isospin unbalance, here called Asymmetric Nuclear Matter (ANM). Other relevant quantities can be easily derived from it, including SNM and PNM.
It is convenient to define an isospin asymmetry parameter as

\begin{eqnarray}
Y=\frac{n_{\text{n}}-n_{\text{p}}}{n}=1-2Y_{p}\;,
\end{eqnarray}

\noindent where $n_{\text{n(p)}}$ is the neutron (proton) density, $n=n_{\text{n}}+n_{\text{p}}$ is the total density of the system and $Y_{p}$ is the proton fraction.
When the asymmetry parameter is equal to $Y=0$ we are in the SNM case, while for $Y=1$ we are in the other extreme case, $i.e.$ PNM.
To present the expressions in a compact, yet transparent, form we define the coefficients $a= \left(3 \pi^2 /2 \right)^{1/3}$ and $b= \left(3 \pi^2  \right)^{1/3}$, and the following functions of the asymmetry parameter
\begin{eqnarray}
F_x &=& \frac{1}{2} \left[ (1+Y)^x + (1-Y)^x \right]\,, \\
G_x &=& \frac{1}{2} \left[ (1+Y)^x - (1-Y)^x \right]\,.
\end{eqnarray}

The EoS in ANM reads 
\begin{eqnarray}\label{eosanm}
E/A &=&\frac{3}{5}\frac{\hbar^{2}}{2m} a^2 F_{5/3} n^{2/3} +\frac{1}{8} \left[ 3 - (2x^{(0)}_{0}+1) Y^2 \right] t^{(0)}_{0} n 
+\frac{1}{48} \left[ 3 - (2x_{3}+1) Y^2 \right]  t_{3}  n^{\alpha+1} \nonumber\\
&& +\frac{3}{80} a^2 \left[ C^{(2)}_{0} F_{5/3} + C^{(2)}_{1} Y G_{5/3} \right] n^{5/3} 
 +\frac{3}{1120} a^4 \left[ C^{(4)}_{0} \left( 5 F_{7/3} +7 F_{5/3}^2 \right)  
+ C^{(4)}_{1} \left( 5 Y G_{7/3} +7 G_{5/3}^2)\right) \right] n^{7/3} \nonumber\\
&&+\frac{1}{240} a^6 \left[ C^{(6)}_{0} \left( 5 F_{3} +27 F_{7/3} F_{5/3} \right)  
+ C^{(6)}_{1} \left( 5 Y G_{3} +27 G_{7/3} G_{5/3} \right) \right] n^3 \;.
\end{eqnarray}
The constants $C^{(n=2,4,6)}_{i=0,1}$ are the following combinations of Skyrme parameters
\begin{eqnarray}
C^{(n)}_{0}&=& 3t^{(n)}_{1}+(5+4x^{(n)}_{2})t^{(n)}_{2} \,,\\
C^{(n)}_{1}&=& -(2x_{1}^{(n)}+1)t_{1}^{(n)}+(2x^{(n)}_{2}+1)t^{(n)}_{2} \, .
\end{eqnarray}

 \begin{figure}[h]
   \centering
   \includegraphics[angle=0,width=0.5\textwidth]{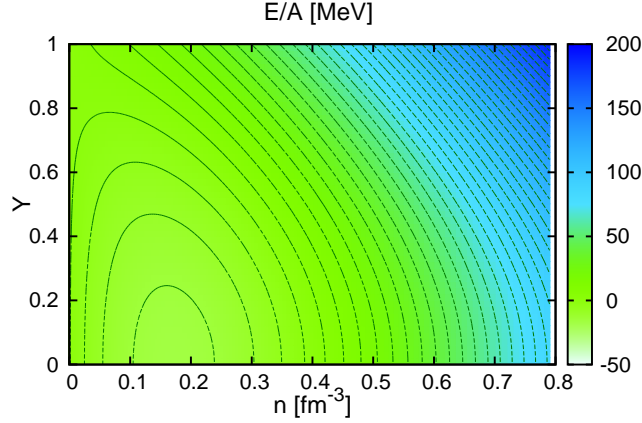}
   \caption{(Colors online) Equations of states in  asymmetric nuclear matter as a function of the density $n$ and asymmetry parameter $Y$.}
              \label{ANM}%
    \end{figure}

In Fig.~\ref{ANM}, we show the binding energy per particle obtained with our new functional as a function of the density and the asymmetry parameter $Y$ of the system. As expected, the energy minimum  is located at $Y=0$ with the values $E/A$ =-17.02 [MeV] and $n_0=0.169$ [fm$^{-3}$]. These values are slightly larger than commonly adopted ones~(\cite{dut12}). This is a drawback of the BBG calculations used to fix the parameters which have $E/A$ =-16.46 [MeV] and $n_0=0.178$ [fm$^{-3}$] (\cite{bal97}).
We decided not to adjust the saturation point to the standard value as extracted, for example, from mass formulas~(\cite{bohr1998nuclear}) and keep the value obtained by the direct fit as done for LNS~(\cite{cao06}).
Due to uncertainties related to three-body forces~(\cite{bal99}) and the methods adopted for the calculations~(\cite{bal12}), these values can change from one \emph{ab initio} method to another. The goal of the present article is to prove that a simple Skyrme functional can grasp the main features of a more complicated calculation based on realistic nucleon-nucleon interaction, as a consequence we prefer not to do any fine tuning around the saturation point since it will not change the main conclusions of the present work.

From  Eq.~\ref{eosanm}, we can also extract other quantities, as the pressure of the system $P=n^2\frac{\partial (E/A)}{\partial n}$ or the nuclear incompressibility $K=9n^2\left.\frac{\partial^2 (E/A)}{\partial n^2}\right|_{n=n_0}=9\left.\frac{\partial P}{\partial n}\right|_{n=n_0}$, as functions of the asymmetry parameter $Y$.  In Tab.\ref{tab:n3lo:properties}, we report several relevant SNM quantities calculated at saturation density. 
Our parametrization gives a value of the incompressibility of $K=231$ MeV at saturation density, which is within the range of acceptable values as discussed by ~\cite{dut12}. The third derivative of the EoS gives us the skewness $Q$.

 \begin{table}[h]
\begin{center}
\caption{Basic SNM properties calculated with the LYVA1 parametrization given in Tab.\ref{tab:n3lo}, the BCPM and the BSk19-21 functionals at saturation density $n_0$.}
\begin{tabular}{c|cccccc}
\hline
\hline
&LYVA1& BCPM & BSk19 & BSk20 & BSk21 & LNS \\
\hline
\hline
$n_0$[fm$^{-3}$] &0.169 &0.160 &0.160 &0.160 &0.158 & 0.175 \\
$E/A$[MeV] &-17.02 & -16.00&-16.08 &-16.80 &-16.05 & -15.31\\
$ K$[MeV] &231 & 214& 237&241 &246& 211\\
$ m^*/m$ & 0.707&1& 0.80& 0.80&0.80&0.825\\
$ Q$[MeV] &-463 &-881 &-298 &-282 &-274&-384\\
$ J$[MeV] & 33.8 &31.9 &30.0 &30.0 &30.0&33.4\\
$ L$[MeV] &64.5 & 53.0 &31.9 &37.4 &46.6&61.5\\
 $K_{sym}$[MeV]&-75.6& -98.1& -191.4&-136.5 &-37.2&-127.7\\
$ Q_{sym}$[MeV]& 464 &877&473 &550 &710& 303\\
  \hline
 \hline
\end{tabular}
\label{tab:n3lo:properties}
\end{center}
\end{table}

The two limiting cases of symmetric nuclear matter and pure neutron matter can be immediately obtained from Eq.~\ref{eosanm}.
\begin{eqnarray}
\left.\phantom{\frac{1}{1}} E/A \right|_{SNM} &=& \frac{3}{5}\frac{\hbar^{2}}{2m} a^2 n^{2/3} + \frac{3}{8} t^{(0)}_{0}n + \frac{1}{16} t_{3}n^{\alpha+1}  +\frac{3}{80} a^2 C^{(2)}_{0} n^{5/3} +\frac{9 }{280} a^4 C^{(4)}_{0} n^{7/3}+ \frac{2 }{15} a^6 C^{(6)}_{0} n^3 \;,\\
\left.\phantom{\frac{1}{1}}E/N \right|_{PNM} &=& \frac{3}{5}\frac{\hbar^{2}}{2m} b^2 n_{\text n}^{2/3} + \frac{1}{4} (1-x^{(0)}_0) t^{(0)}_0  n_{\text{n}}
+ \frac{1}{24} (1-x_3) t_3 n_{\text{n}}^{\alpha+1} \nonumber\\
&&
+\frac{3}{80} b^2 \left[ C^{(2)}_{0}+ C^{(2)}_{1}\right]  n_{\text{n}}^{5/3} 
+\frac{9}{280} b^4 \left[ C^{(4)}_{0}+ C^{(4)}_{1}\right] n_{\text{n}}^{7/3}
+\frac{2}{15} b^6 \left[ C^{(6)}_{0}+ C^{(6)}_{1}\right] n_{\text{n}}^3 \;.
\end{eqnarray}

\begin{figure}[h]
   \centering
   \includegraphics[angle=0,width=0.5\textwidth]{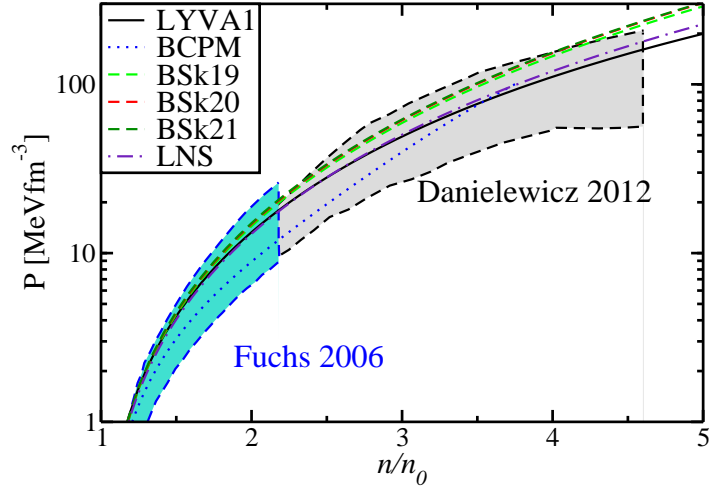}
   \caption{(Colors online) The Pressure as a function of the density in SNM for the different models considered in the article. See text for details.}
              \label{pressure}%
    \end{figure}

In Fig. \ref{pressure}, we show the evolution of the pressure $P$ as a function of the density in SNM for LYVA1, BSk19-21 and BCPM. The two areas represent the constraints obtained on the EoS by \cite{dan02} using experimental observations of heavy-ion collisions and by \cite{Fuc06} of experiments on kaons. We observe that the LYVA1 functional is perfectly consistent with such results.

\begin{figure}[h]
   \centering
   \includegraphics[angle=0,width=0.5\textwidth]{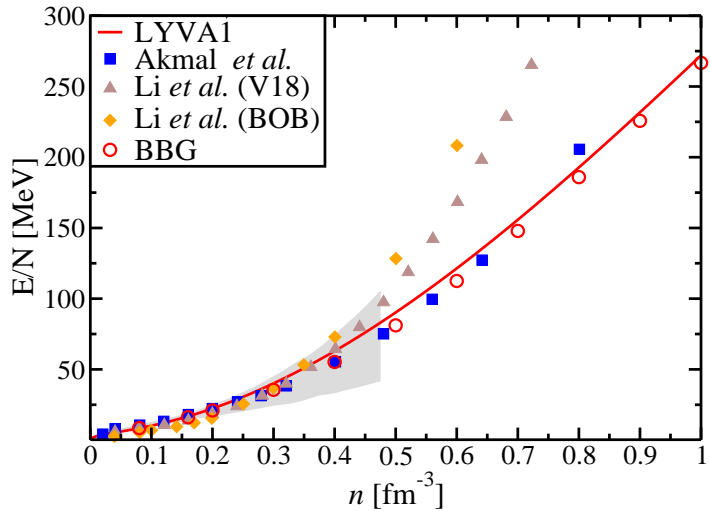}
   \caption{(Colors online) The EoS in PNM for the different models discussed in the text. The shaded area represents the constraints extracted from (\cite{gan12}).}
              \label{eospnm}%
    \end{figure}

In Fig.\ref{eospnm}, we compare the resulting EoS in PNM for the LYVA1 functional and the results of several microscopic calculations (\cite{akm98,li08,bal97,gan12}). We observe that our EoS is compatible with these calculations up to three times the saturation density; beyond that value the different calculations strongly differ from each other. It is thus very important to adopt the same microscopic calculation to constrain both the PNM ans SNM EoS, otherwise, one would obtain non trustable results concerning the behavior of the symmetry energy.

\begin{figure}[h]
   \centering
   \includegraphics[angle=0,width=0.5\textwidth]{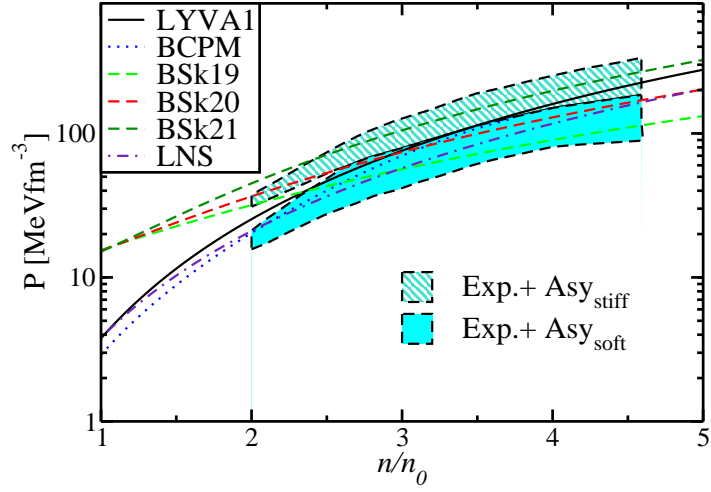}
   \caption{(Colors online) The pressure as a function of the density in PNM for the different models considered in the article. The grey areas are derived from the analysis of (\cite{dan03}).}
              \label{pressurePNM}%
    \end{figure}

In Fig.\ref{pressurePNM}, we show the pressure $P$ in PNM for the different models discussed here and some additional constraints derived in  (\cite{dan03}) and based on the analysis of heavy ion collisions.
We observe that both the BCPM and LYVA1 models are in good agreement with the constraints extracted assuming a soft EoS. See (\cite{dan03}) for more details.

The sound velocity $v_s$ of a system is obtained in the non-relativistic limit from the isothermal incompressibility $\partial n / \partial P$.
We refer to (\cite{haensel2007neutron}) for a more detailed discussion. 
The explicit expression for PNM reads
\begin{eqnarray}
m_n c^2 \left(\frac{v_s}{c} \right)^2 &=&\frac{2}{3}\frac{\hbar^{2}}{2m} b^2 n_{\text n}^{2/3} 
+\frac{1}{2} (1-x_0^{(0)})t_0^{(0)} n_{\text n}
+\frac{1}{24}(1-x_3^{(0)})t_3^{(0)}(1+\alpha)(2+\alpha)n_{\text{n}}^{\alpha+1} \nonumber\\
&&+\frac{1}{6}  b^2 \left[ C^{(2)}_{0}+ C^{(2)}_{1}\right] n_{\text{n}}^{5/3}
+\frac{1}{4} b^4 \left[ C^{(4)}_{0}+ C^{(4)}_{1}\right] n_{\text{n}}^{7/3}
+\frac{8}{5} b^6 \left[ C^{(6)}_{0}+ C^{(6)}_{1}\right] n_{\text{n}}^3.
\end{eqnarray}

\begin{figure}[h]
   \centering
   \includegraphics[angle=0,width=0.5\textwidth]{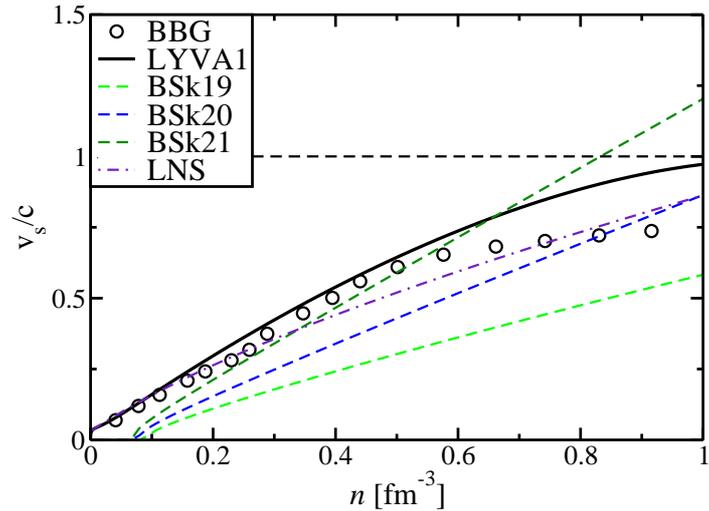}
   \caption{(Colors online) The speed of sound in PNM as a function of the density of the system. The notation is as in Fig.\ref{RESFIT}.}
              \label{vsound}%
    \end{figure}

In Fig.\ref{vsound}, we show the evolution of $v_s$ in PNM as a function of the density. We observe that our parametrization respects the causality principle (there is actually a maximum around 1.1 fm$^{-3}$ with $v_s/c=0.97$), and is thus reliable for the description of high density neutron matter.
On the same figure we also compare the results obtained with the BSk models. We notice that the BSk21 model violates causality in PNM at $n\approx0.8$ fm$^{-3}$. As discussed in (\cite{Gor10B}), the BSk models assure the causality principle in $\beta$-equilibrium nuclear matter for the density ranges found in NS.

From this analysis we exclude the BCPM model since its analytical expressions are strictly valid, by construction, only in the low density regime.

We notice that although the causality principle is always respected by the LYVA1 functional a speed of sound very close to the speed of light is clearly a symptom of the use of the non-relativistic approximation to treat matter very close to the relativisitic limit.


\section{Polarized Matter}\label{Sec:BC}

A major drawback of several effective interactions is the presence of a spontaneous ferromagnetic transition ~(\cite{vid84}) at densities relevant for the physics of nuclei and or NS. However,
such a spontaneous phase transition has not been observed so far by any microscopic calculation~(\cite{pan72,fan01,vid02}). It is thus interesting to determine the behavior of our interaction concerning this aspect.

\begin{figure}[h]
   \centering
   \includegraphics[angle=0,width=0.5\textwidth]{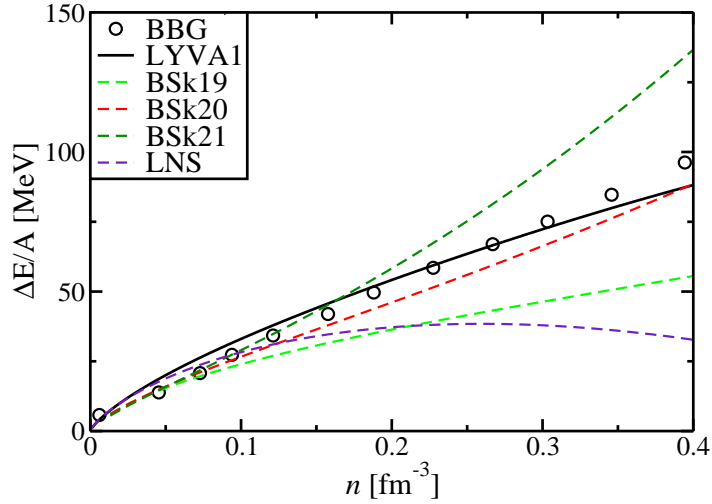}
   \caption{(Colors online) Energy difference between PolPNM and PNM for the different models considered in the text.}
              \label{diffpol}%
    \end{figure}

The expressions for the energy per particle in fully polarized pure neutron matter (PolPNM) reads

\begin{eqnarray}
\left.\phantom{\frac{1}{1}} E/A \right|_{PolPNM} &=&\frac{3}{5}\frac{\hbar^{2}}{2m} c^2  n^{2/3} +\frac{3}{10} c^2(1+x^{(2)}_{2})  t^{(2)}_{2} n^{5/3}+\frac{9}{35} c^4(1+x^{(4)}_{2})  t^{(4)}_{2} n^{7/3}+\frac{16}{15} c^6(1+x^{(6)}_{2})  t^{(6)}_{2} n^{3}
\end{eqnarray}

\noindent where we have used the notation $c=(6\pi^2)^{1/3}$.
In the following we compare the results obtained with the LYVA1 functional and available BBG calculations of~(\cite{bom06}). More precisely, we show in Fig.\ref{diffpol}, the difference of energy per particle between PolPNM and PNM, that is $\Delta E/ A=\left.\phantom{\frac{1}{1}} E/A \right|_{PolPNM}-\left.\phantom{\frac{1}{1}} E/A \right|_{PNM}$ obtained with the LYVA1 functional and the BBG results of ~(\cite{bom06}).
In order to be consistent, we consider only, for this particular case, the results up to $\approx$3 $n_0$ since the treatment of the three-body term at high density is not the same used in BBG results of (\cite{bal97}) and used here for the fit of the LYVA1 functional.
We observe that the LYVA1 as well as BSk20 follow pretty closely the BBG results, while the BSk19(21) tends to underestimate (overestimate) the energy difference between the two systems. The LNS functional is not stable against polarization and at densities $n\approx0.6$ [fm$^{-3}$] favors the appearance of polarized neutron matter.
The BCPM functional has not been included in such analysis since the functional has not been tailored to describe polarized systems.

\section{Symmetry energies}\label{Sec:esym}
 
We give now the LYVA1 expression for the isospin symmetry energy $\varepsilon_{T}(n)$, which plays a crucial role in determining the composition of the NS since the $\beta$-equilibrium condition strongly depends on it. It follows that reproducing the symmetry energy not only at saturation, but also as a function of the density is a necessary condition to have a reliable extrapolation of the high density part of the NS.
Starting from the complete expression of Eq.~\ref{eosanm}, we can expand the binding energy per particle up to second order in the following way
 \begin{eqnarray}
&& \frac{E}{A}(n,Y)= \frac{E}{A}(n,0)+ \varepsilon_{T}(n)Y^2+\dots
\end{eqnarray}
to get the result
\begin{eqnarray}\label{eq:sym}
\varepsilon_{T}(n)=\frac{\hbar^{2}}{6m} a^2 n^{2/3}+\frac{1}{8}C_{1}^{(0)}n 
+\frac{1}{16} a^2 \left\{\frac{1}{3} C_{0}^{(2)}+C_{1}^{(2)} \right\} n^{5/3} 
+\frac{1}{32} a^4 \left\{\frac{4}{3}C_{0}^{(4)}+\frac{8}{3}C_{1}^{(4)} \right\} n^{7/3} 
+ \frac{1}{32} a^6 \left\{\frac{48}{5}C_{0}^{(6)}+16C_{1}^{(6)}\right\} n^3\;.
 \end{eqnarray}

\begin{figure*}[h]
   \centering
   \includegraphics[angle=-90,width=0.4\textwidth]{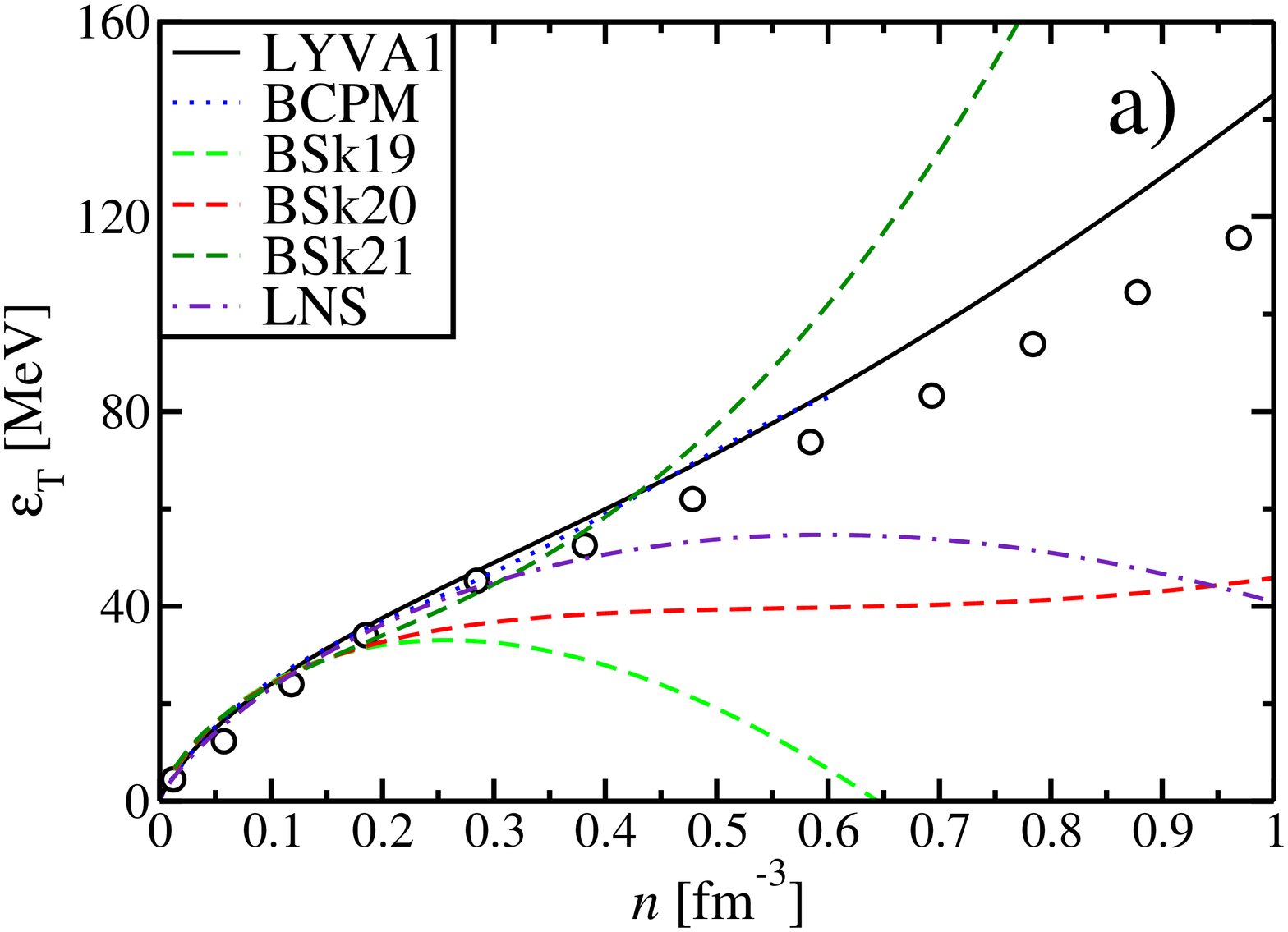}
      \includegraphics[angle=-90,width=0.4\textwidth]{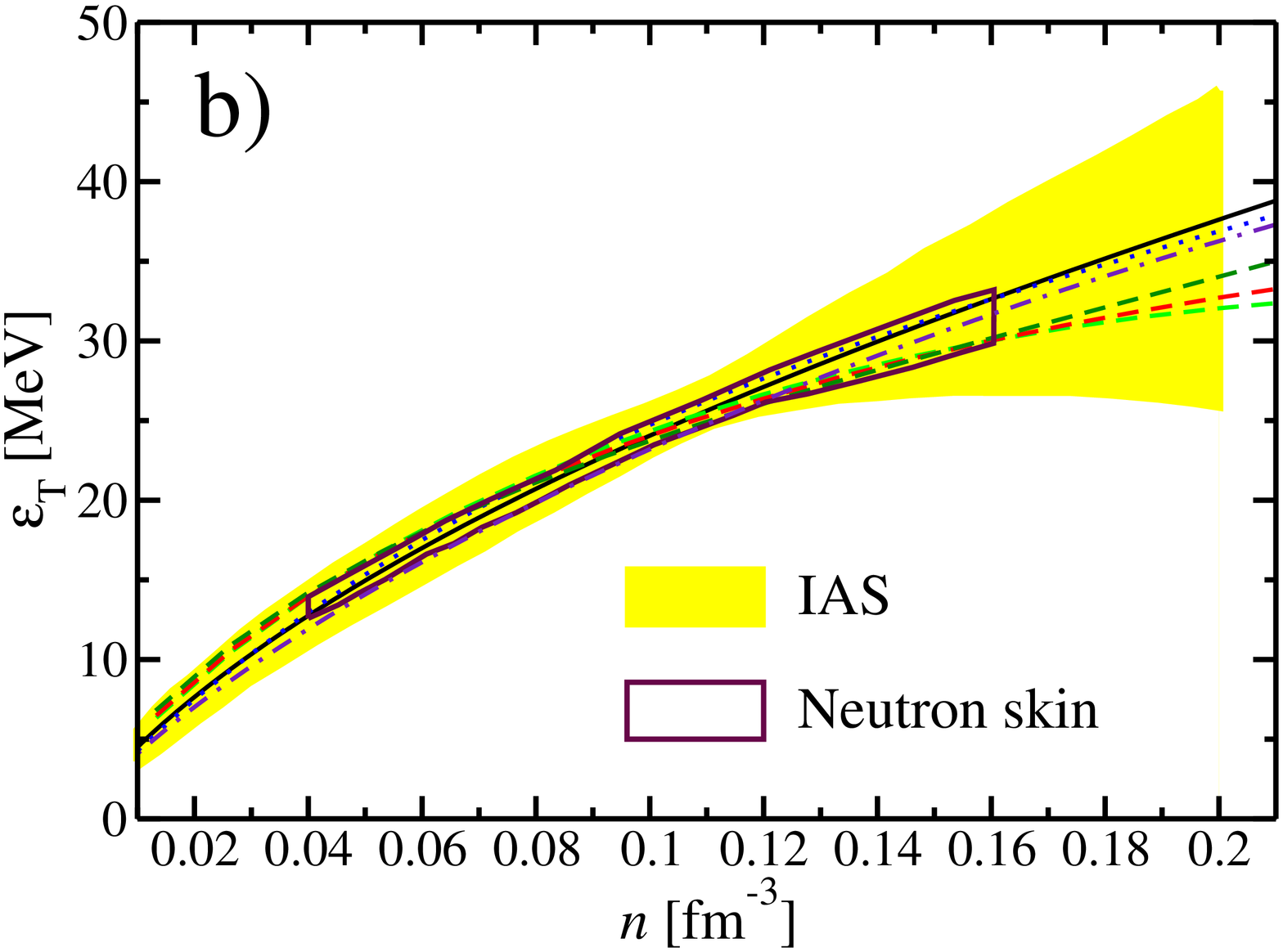}
   \caption{(Colors online) Symmetry energy as a function of the density of the system for different models (left panel). The dots correpond to the BBG calculations. In the right panel  we compare the constraints extracted using IAS ~\cite{dan14}. See text for details.}
              \label{asym}%
    \end{figure*}

In Fig.~\ref{asym}, we show the evolution of the symmetry energy $\varepsilon_T$  as a function of the density of the system. At saturation density, we obtain a value of the symmetry energy $\varepsilon_{T}(n_0)=J=33.8$ MeV, a value compatible with most recent constraints on $J$ obtained combining different experimental data ~(\cite{tsa09}). Furthermore, we observe an excellent agreement up to several times the saturation density value between our results and the BBG ones.
 In the same figure, we also compare the evolution of the symmetry energy for the BCPM and BSk models. We observe that while BCPM gives by construction results which are essentially on top of ours for low density, the BSk give very different behaviors especially beyond saturation density.
There is not agreement between different microscopic approaches concerning the behavior of $\varepsilon_{T}$ beyond saturation density. We refer to the discussion in (~\cite{Gor10B}). 
A possible way to figure out the correct trend of $\varepsilon_{T}$ at high density is the predicted proton fraction and thus the possibility or not of allowing firect URCA process (\cite{han95}). We refer to Sec.\ref{sec:ns}  for a more detailed discussion. We can anyhow anticipate that BSk19-20 and LNS are not compatible with such additional constraints.
In panel (b), we compare the results at low density obtained with the different models and the constraints obtained in (~\cite{dan14}). The large yellow area represents the constraints extracted by analyzing data on isobaric analog states (IAS), while the smaller area delimitated by the solid line contour has been obtained by studying data on the neutron skin properties of some selected nuclei. All the functionals considered in the text respect these constraints.

It is important to remind that  a parabolic approximation was used in (\cite{bal97}) to extract the symmetry energy and the validity of such an approximation has been tested only in the region $n\in[0,0.4]$~fm$^{-3}$~(\cite{bom91,Gor10B}). In this case the symmetry energy is obtained as the difference between the EoS in PNM and SNM

\begin{eqnarray}\label{eq:sym:parab}
\varepsilon_{T}(n)^{(2)}= \left.\frac{E}{A}(n)\right|_{SNM}- \left.\frac{E}{A}(n)\right|_{PNM}\,.
\end{eqnarray}

\begin{figure}[h]
   \centering
   \includegraphics[angle=0,width=0.4\textwidth]{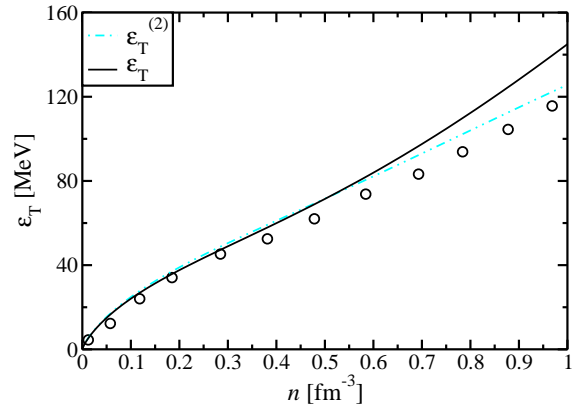}
   \caption{(Colors online) Symmetry energy as a function of the density of the system using different definition of the symmetry energy. The dots represent the BBG results. See text for details.}
              \label{asymparab}%
    \end{figure}

In Fig.\ref{asymparab}, we compare the results for the LYVA1 functional using the definition of Eq.\ref{eq:sym} and Eq.\ref{eq:sym:parab}.
We observe that using the parabolic approximation we have a better reproduction of the high density part of the BBG results of (\cite{bal97}). The agreement is still not perfect due to the small overestimate of the EoS for PNM resulting in our fit and shown in Fig.\ref{RESFIT}. We notice that in our calculations the two definitions of the symmetry energy of Eq.\ref{eq:sym} and Eq.\ref{eq:sym:parab} are essentially on top of each other up to $n\approx0.6$~[fm$^{-3}$].

For completeness, we also define additional quantities related to the symmetry energy, as 
$L= 3n\frac{\partial \varepsilon_{T}(n)}{\partial n }$, 
$K_{sym}= 9n^2\frac{\partial^2 \varepsilon_{T}(n)}{\partial n^2 }$, and
$Q_{sym}= 27 n^3\frac{\partial^3 \varepsilon_{T}(n)}{\partial n^3 }$.
Their values at saturation density are reported in Tab.~\ref{tab:n3lo:properties}. The constraints on these quantities are much less strict, leading to larger error bars~(\cite{dut12}). However, the value obtained here by instance for  $L$ is compatible with some recent estimates~(\cite{che10}) extracted from finite nuclei analysis.  

\section{Effective mass}\label{Sec:effmass}   
 
The effective mass is directly related to some important processes as neutrino emissivity (\cite{yak01}). We thus give the explicit expressions for the neutron ($m^*_n$) and proton ($m^*_p$) Landau effective masses at the Fermi surface as a function of the density and of the asymmetry of the system. For neutrons it reads

\begin{eqnarray}
\frac{\hbar^2}{2m_n^*}-\frac{\hbar^2}{2m_n}&=&\frac{1}{16}\left( C^{(2)}_0 +  C^{(2)}_1 Y \right) n 
+\frac{1}{16} a^2 \left[  \left( C^{(4)}_0   + C_1^{(4)} Y \right) \left( F_{2/3} + G_{2/3} \right)
+ C_0^{(4)} F_{5/3} + C_1^{(4)} G_{5/3} \right] n^{5/3}\;,
\nonumber\\
&+& \frac{1}{16} a^4 \left[ 3 (C_0^{(6)} + C_1^{(6)} Y )( F_{4/3}+G_{4/3}) + 
\frac{42}{5} ( C_0^{(6)} F_{5/3} + C_1^{(6)} G_{5/3} ) (F_{2/3} + G_{2/3} ) 
+ 3 ( C_0^{(6)} F_{7/3} + C_1^{(6)} G_{7/3} ) \right] n^{7/3}\;.
\end{eqnarray}

\noindent The proton effective mass is simply obtained by replacing $Y\rightarrow-Y$ in the previous expression.
It is worth noticing that contrary to the standard Skyrme interaction, the effective mass for our pseudo-potential has an explicit dependence on the momentum $k$, as it happens in the case of real BBG calculations.
See \cite{bec14} for more details.

\begin{figure}[h]
   \centering
   \includegraphics[angle=0,width=0.45\textwidth]{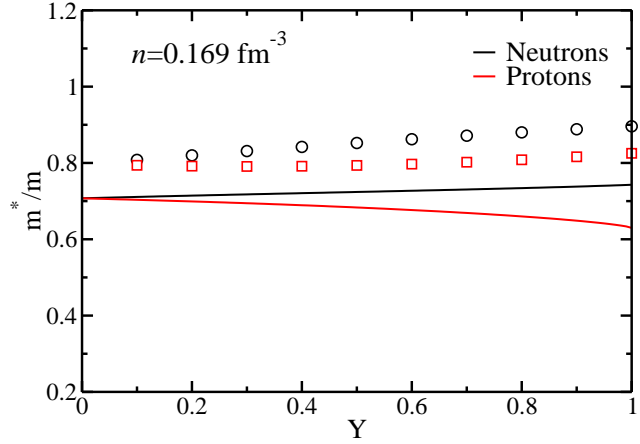}
   \caption{(Colors online) Neutron and proton effective mass at saturation density as a function of the asymmetry parameter $Y$. The dots correspond to the BBG calculations (\cite{bal14}), the solid line to LYVA1 interaction.}
              \label{effmas}%
    \end{figure}

In Fig.~\ref{effmas}, we show the evolution of the effective mass for neutrons and protons as a function of the asymmetry parameter at saturation density. We observe that the mass splitting $\Delta m=m^*_n-m^*_p$ has the correct sign and density behavior as compared to BBG results (\cite{bal14}), although the resulting effective masses are slightly lower at saturation in the SNM case ($m^*/m \simeq 0.7$) compared to the BBG result $(m^*/m \simeq 0.8)$. Such a difference can't be further reduced by a better fine-tuning of the $t_3^{(0)},x_3^{(0)}$ parameters without inducing side effects on other quantities.

\begin{figure*}[h]
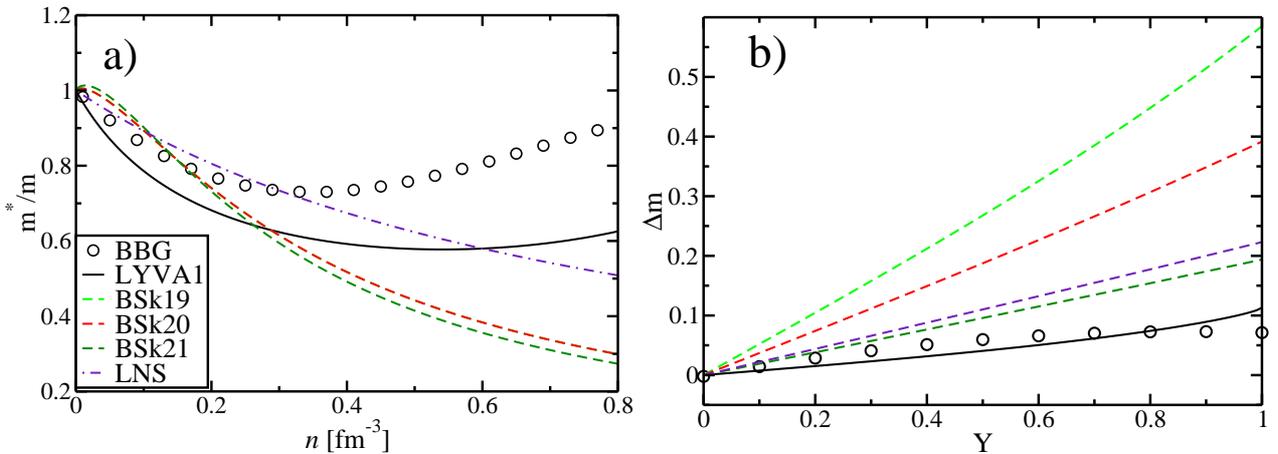

   \centering
   \includegraphics[angle=0,width=0.45\textwidth]{plots/masseffSNM.eps}
     \includegraphics[angle=0,width=0.45\textwidth]{plots/Diff_emass.eps}
   \caption{(Colors online) On the panel a, we show the neutron  effective mass in SNM (Y=0) for the BBG calculations (dots), LYVA1 (solid) and BSk19-21 (dashed). On the panel b the evolution of the effective mass splitting at $n_0=0.169$ fm$^{-3}$  .}
              \label{effmasSNM_PNM}%
    \end{figure*}

In Fig. \ref{effmasSNM_PNM} (a), we compare the effective mass in SNM calculated with the LYVA1 functional and the corresponding BBG results as a function of the density of the system. Although the difference between the two calculations increases with the density reaching at most 30\% in the high density region, the asymptotic behavior is correct with a positive slope at high densities. On the same figure we also show the results obtained with the different BSk models (concerning BCPM, it has been fitted imposing the bare nucleon mass at all densities and all asymmetries).
The major difference between the BSk and LYVA models is related to the high density behavior, where the former lead to a much smaller effective mass compared to BBG results.
 In Fig. \ref{effmasSNM_PNM} (b), we show the evolution of the difference between neutron and proton effective mass $\Delta m= m^*_n/m- m^*_p/m$ at $n_0=0.169$ fm$^{-3}$ for the different models discussed in the present article. We notice that the BSk19-20 models give a much larger splitting than the one observed with BBG calculations.

The presence of magnetic fields inside a star could lead to spin-unbalanced systems. As discussed in Sec.\ref{Sec:BC}, such a configuration does not constitute the ground state of the system up to several times the saturation density, but the presence of an external field could change the situation.
Limiting ourselves to the case of polarized pure neutron matter, it is possible to write the explicit expression for the spin-up ($\uparrow$) and spin-down ($\downarrow$) effective mass as

\begin{eqnarray}\label{effmassup}
\frac{\hbar^2}{2m_{\uparrow}^*}-\frac{\hbar^2}{2m}&=&\frac{1}{8}\left( C^{(2)}_2 +  C^{(2)}_3 \Delta \right) n \\  \nonumber
&+&\frac{1}{8} C^{(4)}_2 n k_F^2 (1+\Delta)^{2/3}+\frac{1}{8} C^{(4)}_3 n k_F^2 \Delta(1+\Delta)^{2/3}\\  \nonumber
&+&\frac{1}{16} C^{(4)}_2 n k_F^2[ (1+\Delta)^{5/3} +(1-\Delta)^{5/3}]+\frac{1}{16} C^{(4)}_3 n k_F^2[ (1+\Delta)^{5/3} -(1-\Delta)^{5/3}]\\ \nonumber
&+&\frac{3}{8} C^{(6)}_2 n k_F^4 (1+\Delta)^{4/3}+\frac{3}{8} C^{(6)}_3 n k_F^4 \Delta(1+\Delta)^{4/3}\\  \nonumber
&+&\frac{21}{40} C^{(6)}_2 n k_F^4(1+\Delta)^{2/3}[ (1+\Delta)^{5/3} +(1-\Delta)^{5/3}]+\frac{21}{40} C^{(6)}_3 n k_F^4(1+\Delta)^{2/3}[ (1+\Delta)^{5/3} -(1-\Delta)^{5/3}]\\  \nonumber
&+&\frac{3}{16} C^{(6)}_2 n k_F^4[ (1+\Delta)^{7/3} +(1-\Delta)^{7/3}]+\frac{3}{16} C^{(6)}_3 n k_F^4[ (1+\Delta)^{7/3} -(1-\Delta)^{7/3}]
\end{eqnarray}

\noindent where we have defined $\Delta=(\rho_{\uparrow}-\rho_{\downarrow})/\rho$. The coupling constants $C^{(n)}_2,C^{(n)}_3$ are related to the $t_i^{(n)},x_i^{(n)}$ as follows

\begin{eqnarray}
C^{(n)}_2&=&(1-x_1^{(n)})t_1^{(n)}+3(1+x_2^{(n)})t_2^{(n)} \\
C^{(n)}_3&=&-(1-x_1^{(n)})t_1^{(n)}+(1+x_2^{(n)})t_2^{(n)} 
\end{eqnarray}

\begin{figure}[h]
   \centering
   \includegraphics[angle=0,width=0.45\textwidth]{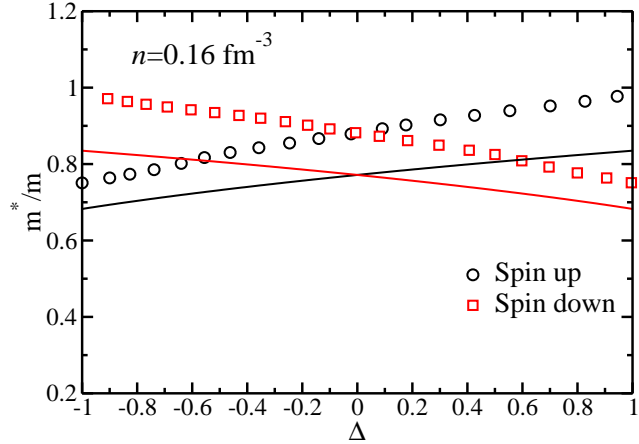}
   \caption{(Colors online) Spin up and spin down neutron effective mass at saturation density as a function of the polarization parameter $\Delta$. The dots correspond to the BBG calculations (\cite{bom06}), the solid line to LYVA1 interaction.}
              \label{effmasSPIN}%
    \end{figure}

\noindent The explicit expressions for the spin-down effective mass can be derived from Eq.\ref{effmassup} by replacing $\Delta\rightarrow-\Delta$.
In Fig. \ref{effmasSPIN}, we show the evolution of the spin up (down) effective mass for the LYVA1 model and the original BBG results (\cite{bom06})

\begin{figure}[h]
   \centering
   \includegraphics[angle=0,width=0.45\textwidth]{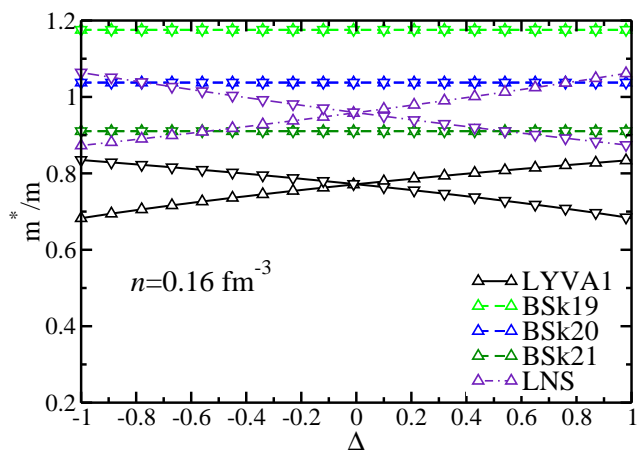}
   \caption{(Colors online) Same as Fig\ref{effmas}, but for other models considered in the text. Up (down) triangle stands for spin up (down) component. }
              \label{effmasBSK}%
    \end{figure}

In Fig.\ref{effmasBSK}, we compare the effective mass for the spin up (down) component for the different functionals considered in the present article in polarized neutron matter. The BCPM results are not present here~: this functional can not be used for polarized systems since the informations on the vector part are missing by construction.  Contrary to the LYVA1 functional, the BSk models do not produce any splitting in the effective mass~: this is due to the absence of terms $C^T\mathbf{s}\cdot \mathbf{T}$ (see Eqs. A1-A2 of \cite{pasgor14}) in the functional which governs such a splitting. The effective mass given by BSk19 is particularly high compared to BBG results.
We refer to (~\cite{Gor10B}) for a more detailed discussion.

\section{Neutron Stars}\label{sec:ns}   
   
\begin{figure}[h]
   \centering
   \includegraphics[angle=0,width=0.5\textwidth]{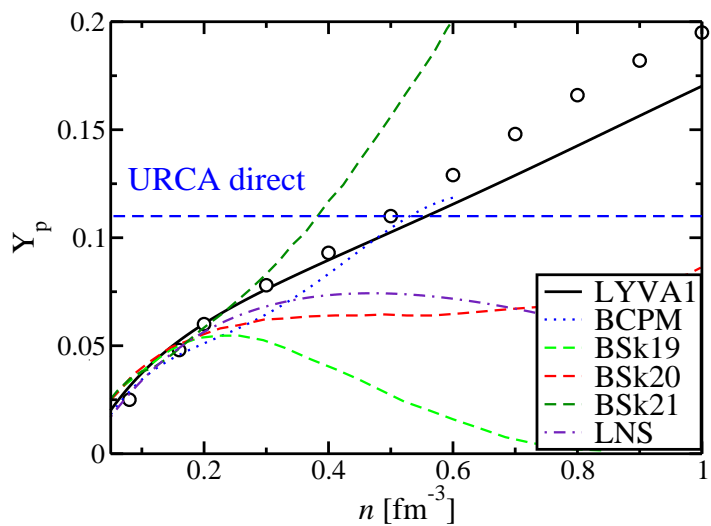}
   \caption{(Colors online) Proton fraction $Y_p$ as a function of the density of the star. The dashed line represents the proton fraction treshold to activate the direct URCA process during the cooling stage of the NS. The open dots represent the BBG calculations.}
              \label{beta}%
    \end{figure}

In this section, we present the basic properties of a non-accreting NS at zero temperature using our interaction. To calculate the mass and the radius of a NS we have to solve the 
Tolman-Oppenheimer-Volkoff (TOV) equations for the total pressure $P$ and the enclosed mass $m$
 
 \begin{eqnarray}\label{eq:tov}
 \frac{dP(r)}{dr}&=&-\frac{G m(r) \varepsilon(r)}{r^{2}}\left[ \left(1+\frac{P(r)\varepsilon(r)}{c^{2}} \right) \left(1+\frac{4\pi r^{3 }P(r)}{\varepsilon(r)c^{2}} \right)\right] \left[ 1-\frac{2G m(r)}{rc^{2}}\right]^{-1}\;,\nonumber\\
 \frac{dm(r)}{dr}&=&4\pi r^{2}  \varepsilon(r)\;,
 \end{eqnarray}  
   
 \noindent where $G$ is the gravitational constant and $ \varepsilon(r)$ is the total energy density of the system. 
Since in our model we consider only neutrons, protons and electrons, the $\beta$-equilibrium condition at each value of the density of the star translates into the equation $\mu_p+\mu_n=\mu_e$ for the chemical potentials, the possible contribution of muons being neglected. In Fig.~\ref{beta}, we show the proton fraction inside the star calculated using our EoS. According to (\cite{han95}), the direct URCA process, which is very important to have a fast cooling during stellar evolution (\cite{lat91}), can take place when the proton fraction is $x\approx0.11$. We observe that our model predicts the possibility for direct URCA process at already 3 times saturation density.
On the same plot, we also present the BBG results. The agreement between the two calculations is very good up to $n \simeq 0.6$~fm$^{-3}$. 

As anticipated in Sec.\ref{Sec:esym}, the possibility of allowing or not a direct URCA process can be used to make some consideration concerning the behavior of the symmetry energy in the high density region. The $\beta$-equilibrium condition for nucleonic equation of state can be related directly, within the parabolic approximation, to the symmetry energy, we have

\begin{eqnarray}
\mu_e=\mu_n-\mu_p\approx 4\varepsilon_T(n)(1-2Y_p)
\end{eqnarray}

where $\mu_{e,np,}$ are the chemical potential of the different species included in the EoS.
From Fig.\ref{beta}, we can conclude that only the LYVA1, BSk21 and BCPM functionals allow for a direct URCA process in NS.

\begin{figure*}[h]
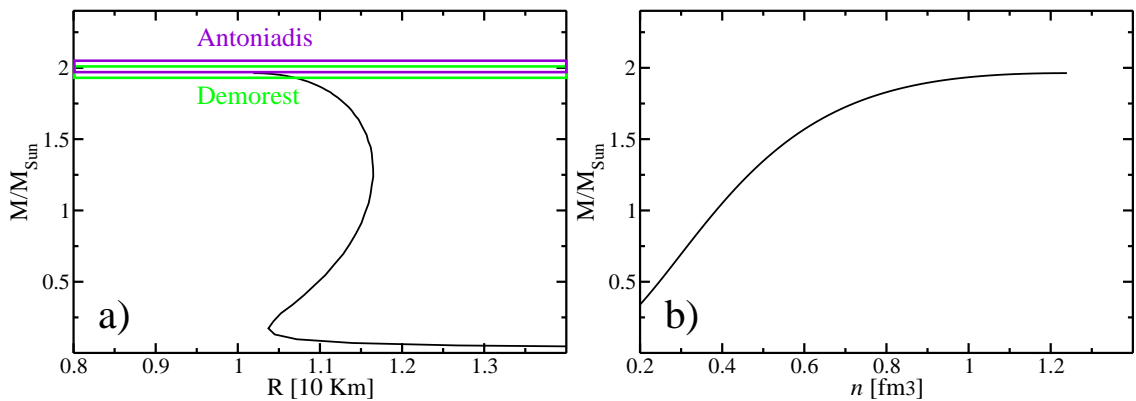

   \centering
   \includegraphics[angle=0,width=0.4\textwidth]{plots/NS_mass.eps}
   \includegraphics[angle=0,width=0.4\textwidth]{plots/NS_massRho.eps}

   \caption{(Colors online) In panel (a), we show the mass-radius relation for NS obtained with our EoS by solving TOV equations. The two horizontal bars refers to the two recent of NS masses measurements $M/M_{Sun}=1.97\pm0.04$ given in ~(\cite{dem10}) and $M/M_{Sun}=2.01\pm0.04$ given in ~(\cite{ant13}). In panel (b), the mass-density relation. }
              \label{ns:tov}%
    \end{figure*}

To describe the structure of the NS we need to solve  Eq. (\ref{eq:tov}) imposing the $\beta$-equilibrium for each value of the density. In the most external layers, the crust, it is possible to observe the presence of structures, either nuclei or more exotic $pasta$-phases (\cite{cha08}). This part of the star will be described by using the EoS of  (\cite{dou11}), which has been derived using the SLy4 functional and by means of the  Compressible Liquid Drop Model~(\cite{dou00}). It allows for a simple description of both the crust (inner and outer), but also for the liquid-core transition. We match our EoS with the one of (\cite{dou11}) at $n\approx0.08$ fm$^{-3}$. This small inconsistency in the EoS will not affect the value of the maximum mass of the star, but it introduces an error of at most 5\% on its radius~(\cite{heb10}). The study of the inhomogeneous phase of the NS with our functional will be the subject of a forthcoming study. Our results concerning the Mass-Radius relation are shown in Fig.~\ref{ns:tov}(a).
 %
%
On the same plot, we also show the recent measurements of masses of NS (\cite{dem10,ant13}), which are both compatible for a $2M_{Sun}$ neutron star. We observe that our EoS gives a prediction compatible with the latest experimental measurements giving a maximal value of $M=1.96 M_{Sun}$ in the non-rotating case. 
The inclusion of extra degrees of freedom as pions, kaons or hyperons would affect that result (\cite{Hei00}), as well as the effect of rotation (\cite{sal94,ste03}). In view of the results of (\cite{sal94b}), we could expect an increase of $\approx10-15$\% for the results of our EoS. The detailed study of these effects goes beyond the scope of this paper and we leave it for the future.

The radius of a NS is very difficult to extract from observations due to the several hypothesis one has to do on the atmosphere composition. Different models (\cite{sul11,ste10}) lead to slightly different values for the radius, but it is nevertheless possible to give an upper value of around 12.5 km for a NS with a mass $1.4M_{Sun}$. See also discussion in (\cite{for14}). From the EoS of our functional, we get 11.6 km, in fair agreement with the original BBG results. In  Fig.~\ref{ns:tov}(b), we show the evolution of the maximum mass of the NS as a function of the central density of the star. It is worth noticing that recent constraints of (\cite{kla06}) implies the absence of direct URCA process for NS within a mass range of $1-1.5M_{Sun}$. From Fig.\ref{ns:tov} (b), we can observe that the lowest value of the density at which URCA process take place correspond to a NS of mass $1.54\;M_{Sun}$.

\section{Conclusions}\label{sec:concl} 

We have presented a new nucleonic Equation of State based on the \emph{extended} Skyrme functional. The inclusion of higher order derivative terms allow us to give a more precise description of the high-density region.
By fixing the coupling constants of our functional on the EoS of {\em ab initio} calculations, we have shown the possibility of extracting analytically several quantities of strong astrophysical interest as incompressibility, pressure or effective mass. As shown in (\cite{dav15}), the higher order gradients allow us to grasp the correct physical behavior obtained with a microscopic calculation. Our functional can be fitted to other microscopic calculations, thus providing a more powerful tool for astrophysics than a simple interpolation procedure. The  Skyrme functional can be easily implemented to perform calculations in all layers of the star, not only in its uniform phase (\cite{dou11,pea12}).

We have compared our results with other commonly adopted functionals, namely the BCPM  (\cite{bal13,bal14NS,sha15}) and the BSk functionals (~\cite{gor09,cha10,Gor10B,gor13}).
We have shown that our model is complementary to the results obtained to these different group since it aims at reproducing as accurately as possible BBG results in nuclear matter (including polarized matter), so that the functional can be used to describe both ground state properties and excited states.

Due to the simplicity of the calculations, the formalism can be easily extended to properly include finite-temperature effects. In that case, it is necessary to replace the step function used to evaluate the EoS of the system by a Fermi-Dirac distribution~(\cite{bon81}). These integrals can be approximated with analytical expressions as shown by \cite{ant93}. The possible temperature dependence of the coupling constants can be also studied by direct comparison with existing microscopic calculations at finite temperature ~(\cite{pan89,lej86}).  


\begin{acknowledgements}
We thank M. Oertel for enlightening discussions and a careful reading of the manuscript. We thank N. Chamel for interesting suggestions to the manuscript.  We thank M. Baldo and K. Hebeler for providing us
with the results of their calculations. JN has been supported by grant FIS2014-51948-C2-1-P, Mineco (Spain).
\end{acknowledgements}

\bibliographystyle{aa} 
\bibliography{biblio}
\end{document}